\documentclass[pra,twocolumn,tightenlines,nofootinbib]{revtex4}
\usepackage{bm,dcolumn,amsmath,graphicx}

\begin{document}
\title{
Magic wavelengths,  matrix elements, polarizabilities, and lifetimes of Cs
}

\author{M. S. Safronova$^{1,2}$}
\author{U. I. Safronova$^{3,4}$}
\author{Charles W. Clark$^{2}$}
\affiliation{}

\affiliation {$^1$Department of Physics and Astronomy, 217 Sharp
Lab, University of Delaware, Newark, Delaware 19716\\
 $^2$Joint Quantum Institute, National Institute of Standards and Technology
and the University of Maryland, College Park, Maryland 20742\\
$^3$Physics Department, University of Nevada, Reno, Nevada 89557\\
\\$^4$Department of Physics,  University of Notre Dame,
Notre Dame, IN 46556
}

\date{\today}

\begin{abstract}
Motivated by recent interest in their applications, we report a systematic study of Cs atomic properties calculated by a
high-precision relativistic all-order method. Excitation energies,
reduced matrix elements, transition rates,
and lifetimes are determined for  levels with principal quantum numbers  $n \leq 12$
and orbital angular momentum quantum numbers $l \leq 3$.
Recommended values and estimates of uncertainties are
provided for a
 number of electric-dipole transitions and the electric dipole
polarizabilities of the $ns$, $np$, and $nd$ states. We also report a
calculation
of the electric quadrupole polarizability of the ground state.
We display the dynamic polarizabilities of the $6s$ and $7p$ states
for optical wavelengths between 1160~nm and 1800~nm and identify corresponding
magic wavelengths for the
  $6s-7p_{1/2}$, $6s-7p_{3/2}$
transitions. The values of relevant matrix elements needed
for polarizability calculations at other wavelengths are
 provided.
\end{abstract}
\maketitle

\section{Introduction}

Cs atoms are used in a wide range of applications including atomic clocks and realization of the second \cite{HeaDonLev14,JefHeaPar14}, the most precise low-energy test of the Standard Model of the electroweak
interactions \cite{PorBelDer09},
search for spatiotemporal variation of fundamental constants \cite{HunLipTam14},
study of degenerate quantum gases \cite{ClaHaXu15}, qubits of
quantum information systems \cite{WanZhaCor15},
search for the  electric-dipole moment of the electron \cite{TanZhaWei15},
atom interferometry \cite{HamJafBro15},
atomic magnetometry \cite{PatZhiHov14}, and
tests of Lorentz invariance \cite{WolChaBiz06}.
As a result, accurate knowledge of Cs atomic properties, in particular electric-dipole matrix elements, lifetimes, polarizabilities, and magic wavelengths became increasingly important.
While a number of $6s-np$ and $7s-np$ transitions have been studied in detail owing to their importance to test of the Standard Model \cite{PorBelDer09}, recent applications require reliable values of many other properties.

Many of the applications listed above involve optically trapped Cs atoms.
The energy levels of atoms trapped in a light field
are generally shifted by a quantity that is proportional to their frequency-dependent polarizability \cite{mitroy-10}.
It is often beneficial to minimize the resulting ac Stark shift of transitions between different levels, for example
in cooling or trapping applications.
At a ``magic wavelength'', which was first used in atomic clock
applications \cite{KatIdoKuw99,YeVerKim99}, the ac Stark shift of a transition is zero.
Magic wavelengths for $6s-7p$ transitions are of potential use for state-insensitive cooling and trapping
and have not been previously calculated. Similar $2s-3p$ and $4s-5p$ transitions have been recently used in $^6$Li \cite{DuaHarHit11} and $^{40}$K \cite{McKJerFin11}, respectively, as alternatives to conventional cooling with the $2s-2p$ and $4s-4p$ transitions.

Here we report an extensive study of a variety of Cs properties of experimental interest. We use several variants of the relativistic high-precision all-order  (linearized
 coupled-cluster) method \cite{SafJoh08}
to critically evaluate the accuracy of our calculations and provide recommended values with associated uncertainties.
 Atomic properties of Cs are evaluated
for $ns$, $np$, $nd$, and $nf$
 states with $n \leq 12$ . Excitation energies and
lifetimes are calculated for the lowest 53 excited states.  The
reduced electric-dipole matrix elements, line strengths,  and
transition rates are determined for allowed transitions between
these levels.   The static electric quadrupole polarizability is
determined for the $6s$ level. Scalar and tensor polarizabilities
of  $(5-9)d$, $(6-9)p$, and $(7-10)s$  states of Cs are evaluated. The uncertainties of the
final values are estimated for all properties.

As a result of these calculations, we are able to identify the magic wavelengths for the
  $6s-7p_{1/2}$ and $6s-7p_{3/2}$
transitions in the 1160~nm and 1800~nm wavelength range.

\begin{table*}
\caption{\label{table1} Recommended values of the reduced
electric-dipole matrix elements $D$ in cesium in atomic units. Uncertainties are given in
parenthesis.
Absolute values are given.}
\begin{ruledtabular}
\begin{tabular}{lrlrlrlrlr}
\multicolumn{1}{c}{Transition}& \multicolumn{1}{c}{D}&
\multicolumn{1}{c}{Transition}& \multicolumn{1}{c}{D}&
\multicolumn{1}{c}{Transition}& \multicolumn{1}{c}{D}&
\multicolumn{1}{c}{Transition}& \multicolumn{1}{c}{D}&
\multicolumn{1}{c}{Transition}& \multicolumn{1}{c}{D}\\
\hline
$ 7s   - 6p_{1/2}$&  4.24(1)     &  $ 6p_{1/2} -  7d_{3/2}$&    2.05(2) &$ 5d_{3/2}- 6p_{1/2}$&   7.1(1) &  $ 6d_{3/2} -  6p_{1/2}$&    4.23(7) &  $ 5d_{3/2} -  7f_{5/2}$&    1.73(1) \\
$ 7s   - 6p_{3/2}$&  6.48(2)     &  $ 6p_{3/2} -  7d_{3/2}$&   0.976(0) &$ 5d_{3/2}- 7p_{1/2}$&   2.3(4) &  $ 6d_{3/2} -  7p_{1/2}$&   17.99(4) &  $ 5d_{3/2} -  8f_{5/2}$&    1.34(1) \\
$ 7s   - 7p_{1/2}$& 10.31(4)     &  $ 6p_{3/2} -  7d_{5/2}$&    2.89(3) &$ 5d_{3/2}- 8p_{1/2}$&  0.63(5) &  $ 6d_{3/2} -  8p_{1/2}$&     5.0(2) &  $ 5d_{5/2} -  6f_{5/2}$&   0.644(6) \\
$ 7s   - 7p_{3/2}$& 14.32(6)     &  $ 6p_{3/2} - 10d_{5/2}$&   0.979(6) &$ 5d_{3/2}- 9p_{1/2}$&  0.34(2) &  $ 6d_{3/2} -  9p_{1/2}$&    1.56(2) &  $ 5d_{5/2} -  6f_{7/2}$&    2.88(3) \\
$ 7s  -  8p_{1/2}$&  0.914(27)    &  $ 6p_{3/2} - 11d_{5/2}$&   0.782(5) &$ 5d_{3/2}-10p_{1/2}$&  0.22(1) &  $ 6d_{3/2} - 10p_{1/2}$&    0.85(2) &  $ 5d_{5/2} -  7f_{5/2}$&   0.468(3) \\
$ 7s  -  8p_{3/2}$&  1.620(35)    &  $ 7p_{1/2} -  6d_{3/2}$&   17.99(4) &$ 5d_{3/2}-11p_{1/2}$& 0.164(9) &  $ 6d_{3/2} - 11p_{1/2}$&    0.56(1) &  $ 5d_{5/2} -  7f_{7/2}$&    2.10(1) \\
$ 7s  -  9p_{1/2}$&  0.349(10)    &  $ 7p_{3/2} -  6d_{3/2}$&    8.07(2) &$ 5d_{3/2}-12p_{1/2}$& 0.128(7) &  $ 6d_{3/2} - 12p_{1/2}$&   0.415(8) &  $ 6d_{3/2} -  4f_{5/2}$&   24.62(9) \\
$ 7s  -  9p_{3/2}$&  0.680(14)    &  $ 7p_{3/2} -  6d_{5/2}$&   24.35(6) &$ 5d_{3/2}- 6p_{3/2}$&  3.19(7) &  $ 6d_{3/2} -  6p_{3/2}$&    2.09(3) &  $ 6d_{5/2} -  4f_{5/2}$&    6.60(2) \\
$ 7s  - 10p_{1/2}$&  0.191(6)     &  $ 7p_{3/2} - 10d_{3/2}$&   0.680(6) &$ 5d_{3/2}- 7p_{3/2}$&   0.9(2) &  $ 6d_{3/2} -  7p_{3/2}$&    8.07(2) &  $ 6d_{5/2} -  4f_{7/2}$&    29.5(1) \\
$ 7s  - 10p_{3/2}$&  0.396(9)     &  $ 7p_{3/2} - 10d_{5/2}$&    2.02(2) &$ 5d_{3/2}- 8p_{3/2}$&  0.26(3) &  $ 6d_{3/2} -  8p_{3/2}$&    1.98(7) &  $ 7d_{3/2} -  5f_{5/2}$&    43.4(2) \\
$ 7s  - 11p_{1/2}$&  0.125(4)     &  $ 7p_{3/2} - 11d_{5/2}$&    1.55(1) &$ 5d_{3/2}- 9p_{3/2}$&  0.14(1) &  $ 6d_{3/2} -  9p_{3/2}$&   0.633(9) &  $ 7d_{5/2} -  5f_{5/2}$&   11.66(5) \\
$ 7s  - 11p_{3/2}$&  0.270(7      &  $ 8p_{1/2} -  7d_{3/2}$&    32.0(1) &$ 5d_{3/2}-10p_{3/2}$& 0.091(7) &  $ 6d_{3/2} - 10p_{3/2}$&   0.346(7) &  $ 7d_{5/2} -  5f_{7/2}$&    52.2(2) \\
  $ 8s   -  7p_{1/2}$&    9.31(2)  &  $ 8p_{3/2} -  7d_{3/2}$&   14.35(5) &$ 5d_{3/2}-11p_{3/2}$& 0.067(5) &  $ 6d_{3/2} - 11p_{3/2}$&   0.230(5) &  $ 8d_{3/2} -  6f_{5/2}$&    65.2(5) \\
  $ 8s   -  7p_{3/2}$&   14.07(7)  &  $ 8p_{3/2} -  7d_{5/2}$&    43.2(1) &$ 5d_{3/2}-12p_{3/2}$& 0.052(4) &  $ 6d_{3/2} - 12p_{3/2}$&   0.169(4) &  $ 8d_{5/2} -  6f_{5/2}$&    17.5(1) \\
  $ 8s   -  8p_{1/2}$&   17.78(6)  &  $ 9p_{1/2} -  8d_{3/2}$&    49.3(1) &$ 5d_{5/2}- 6p_{3/2}$&   9.7(2) &  $ 6d_{5/2} -  6p_{3/2}$&    6.13(9) &  $ 8d_{5/2} -  6f_{7/2}$&    78.4(6) \\
  $ 8s   -  8p_{3/2}$&   24.56(9)  &  $ 9p_{3/2} -  8d_{3/2}$&   22.14(7) &$ 5d_{5/2}- 7p_{3/2}$&   2.8(5) &  $ 6d_{5/2} -  7p_{3/2}$&   24.35(6) &  $ 9d_{3/2} -  7f_{5/2}$&    90.5(9) \\
  $ 9s   -  7p_{1/2}$&    1.96(2)  &  $ 9p_{3/2} -  8d_{5/2}$&    66.6(2) &$ 5d_{5/2}- 8p_{3/2}$&  0.80(7) &  $ 6d_{5/2} -  8p_{3/2}$&     6.2(2) &  $ 9d_{5/2} -  7f_{5/2}$&    24.4(2) \\
  $ 9s   -  8p_{1/2}$&   16.06(4)&  $10p_{1/2} -  9d_{3/2}$&    70.0(1) &$ 5d_{5/2}- 9p_{3/2}$&  0.43(3) &  $ 6d_{5/2} -  9p_{3/2}$&    1.97(3) &  $ 9d_{5/2} -  7f_{7/2}$&   108.9(0) \\
  $ 9s   -  8p_{3/2}$&   24.13(5)&  $10p_{3/2} -  9d_{3/2}$&   31.45(8) &$ 5d_{5/2}-10p_{3/2}$&  0.28(2) &  $ 6d_{5/2} - 10p_{3/2}$&    1.07(2) &  $10d_{5/2} -  6f_{5/2}$&    2.14(2) \\
  $ 9s   -  9p_{1/2}$&   27.10(8)&  $10p_{3/2} -  9d_{5/2}$&    94.5(2) &$ 5d_{5/2}-11p_{3/2}$&  0.21(1) &  $ 6d_{5/2} - 11p_{3/2}$&    0.71(1) &  $10d_{5/2} -  6f_{7/2}$&    9.56(8) \\
  $ 9s   -  9p_{3/2}$&    37.3(1)&  $11p_{1/2} - 10d_{3/2}$&    94.1(2) &$ 5d_{5/2}-12p_{3/2}$&  0.16(1) &  $ 6d_{5/2} - 12p_{3/2}$&    0.53(1) &  $10d_{5/2} -  8f_{5/2}$&    32.1(3) \\
  $10s   -  7p_{1/2}$&   0.999(9)&  $11p_{3/2} - 10d_{3/2}$&   42.29(9) &$ 7d_{3/2} - 6p_{1/2}$&  2.05(2)&  $ 7d_{5/2} -  6p_{3/2}$&    2.89(3) &  $10d_{5/2} -  8f_{7/2}$&    143(1)  \\
  $10s   -  8p_{1/2}$&    3.15(3)&  $11p_{3/2} - 10d_{5/2}$&   127.0(2) &$ 7d_{3/2} - 7p_{1/2}$&   6.6(2)&  $ 7d_{5/2} -  7p_{3/2}$&     9.6(3) &  $11d_{5/2} -  7f_{5/2}$&    3.08(3) \\
  $10s   -  9p_{1/2}$&   24.50(5)&  $12p_{1/2} - 11d_{3/2}$&   121.6(2) &$ 7d_{3/2} - 8p_{1/2}$&  32.0(1)&  $ 7d_{5/2} -  8p_{3/2}$&    43.2(1) &  $11d_{5/2} -  7f_{7/2}$&    13.8(1) \\
  $10s   -  9p_{3/2}$&   36.69(8)&  $12p_{3/2} - 11d_{3/2}$&   54.66(9) &$ 7d_{3/2} - 9p_{1/2}$&   9.0(2)&  $ 7d_{5/2} -  9p_{3/2}$&    11.1(2) &  $12d_{5/2} -  8f_{5/2}$&    4.21(4) \\
  $11s   -  7p_{1/2}$&   0.650(6) & $12p_{3/2} - 11d_{5/2}$&   164.1(2) &$ 7d_{3/2} - 9p_{3/2}$&  3.56(9)&  $ 7d_{5/2} - 10p_{3/2}$&     3.6(1) &  $12d_{5/2} -  8f_{7/2}$&    18.8(2) \\
 $11s   -  8p_{1/2}$&    1.56(1) & $12s   -  8p_{1/2}$&   1.002(9)&    $7d_{3/2} -  6p_{3/2}$& 0.976(0)&      $ 7d_{3/2} - 12p_{3/2}$&  0.42(1) &   $ 7d_{3/2} -  8p_{3/2}$& 14.35(5)     \\
 $11s   -  9p_{1/2}$&    4.61(4) & $12s   -  9p_{1/2}$&    2.24(2)&    $7d_{3/2} -  7p_{3/2}$&   3.3(1)&      $ 7d_{3/2} - 10p_{1/2}$&  2.86(8) &   $ 7d_{5/2} - 11p_{3/2}$&  1.97(6)     \\ $11s   -  9p_{3/2}$&    6.29(6) &  $13s   -  7p_{1/2}$&   0.370(3) &  $ 7d_{3/2} - 10p_{3/2}$&  1.16(4) &   $ 7d_{3/2} - 11p_{1/2}$&  1.55(4)&    &   \\
 $12s   -  7p_{1/2}$&   0.474(5) &  $13s   -  8p_{1/2}$&   0.727(6) &  $ 7d_{3/2} - 11p_{3/2}$&  0.64(2) &   $ 7d_{3/2} - 12p_{1/2}$&  1.03(3)&    &
\end{tabular}
\end{ruledtabular}
\end{table*}
\section{Previous Cs polarizability studies}

In 2010, Mitroy {\it et al.\/} \cite{mitroy-10} reviewed the theory and applications of atomic and ionic polarizabilities across the periodic table of the elements. The static and dynamic polarizabilities of Cs have increased in interest recently, as demonstrated by a number of
experimental
\cite{pol-14,pol-11a,pol-11b,pol-11c,hyp-07,pol-07a,pol-07b,tran-04,pol-03a,pol-03c,pol-99a,pol-98a,pol-97b}
and theoretical
\cite{pol-14a,pol-14b,pol-13a,pol-10a,pol-10b,mitroy-10,pol-10d,pol-09a,pol-08a,pol-cs-07,pol-06a,pol-06b,pol-06c,pol-05a,pol-04a,pol-02a,pol-99b,pol-99c,pol-97a,pol-94a,pol-93a}
studies.

Safronova {\it et al.\/} \cite{pol-06c} presented results of
first-principles calculations of the frequency-dependent
polarizabilities of all alkali-metal atoms for light in the
wavelength range 300-1600~nm, with particular attention to
wavelengths of common infrared lasers.
High-precision study of Cs polarizabilities for a number of states was presented in
Ref.~\cite{pol-cs-07}.
Inconsistencies between $5d$ lifetimes and $6p$  polarizability measurements
in Cs was investigated by Safronova and Clark \cite{pol-04a}.
The {\it ab initio} calculation of $6p$ polarizabilities were found to agree
with experimental values  \cite{pol-04a}.
An experimental and theoretical study of the $6d_{3/2}$
polarizability of cesium was reported by Kortyna {\it et al.\/}
\cite{pol-11b}. The scalar and tensor polarizabilities were
determined from hyperfine-resolved Stark--shift measurements using
two-photon laser-induced-fluorescence spectroscopy of an effusive
beam.
Auzinsh {\it et al.\/} \cite{hyp-07} presented an experimental and
theoretical investigation of the polarizabilities and hyperfine
constants of $nd$ states in $^{133}$Cs.
Experimental values for the hyperfine constant $A$ were obtained
from level-crossing signals of the $(7,9,10)d_{5/2}$ states of
Cs and precise calculations of the tensor polarizabilities
 $\alpha_2$. The results of relativistic many-body calculations for scalar
and tensor polarizabilities of the $(5-10)d_{3/2}$ and
$(5-10)d_{5/2}$ states were presented and compared with measured
values from the literature.
Gunawardena  {\it et al.\/} \cite{pol-07b} presented results of a
precise determination of the static polarizability of the $8s$ state of atomic cesium, carried out jointly through
experimental measurements of the dc Stark shift of the $6s \rightarrow  8s$  transition using
Doppler-free two-photon absorption and theoretical computations
based on a relativistic all-order method.

\section{Electric-dipole matrix elements and lifetimes of cesium}
\label{Section1}
We carried out several calculations using different methods of
increasing accuracy: the lowest-order Dirac-Fock approximation (DF), second-order relativistic
many-body perturbation theory (RMBPT), third-order RMBPT, and several variants of the
linearized coupled-cluster (all-order) method. Comparing
values obtained in different approximations allows us to evaluate the size of the
 higher-order correlations corrections beyond the third
order and estimate some omitted classes of the high-order correlations correction.
As a result, we can present recommended values of Cs properties and estimate their uncertainties.
The RMBPT calculations are carried out using the
method described in Ref.~\cite{adndt-96}. A review of the all-order method, which involves summing
series of dominant many-body perturbation terms to all orders,
is given in \cite{SafJoh08}. In the single-double (SD) all-order approach, single and double excitations of the
Dirac-Fock orbitals are included. The SDpT all-order approach also includes classes of the triple
excitations. Omitted higher excitations can also be estimated by the scaling procedure described in
\cite{SafJoh08}, which can be started from either SD or SDpT approximations. We carry out all four of such all-order
computations, \textit{ab initio} SD and SDpT and scaled SD and SDpT.

The removal energies for a large number of Cs states, calculated in various approximations are
given in Table I of the Supplemental Material \cite{SM}. The accuracy of the energy levels is a
good general indicator of the overall accuracy of the method.
The all-order {\textit{ab initio}}  SD and SDpT values for the ground state ionization potential differ from the experiment \cite{nist-web} by 0.4\% and 0.58\% respectively. Final {\it{ab initio}} SDpT all-order energies are in excellent agreement with experiment, to 0.05-0.4\% for all levels with the exception of the $5d$ states, where the difference is 1.3\%-1.4\%.
The larger discrepancy with experiment for the $5d$ states is explained by significantly larger correlation corrections, 16\% for the $5d_{3/2}$ state in comparison with only 7\% for the $6p_{1/2}$ state. In the isoelectronic spectra of Ba$^+$ and La$^{2+}$, on the other hand, the correlation corrections of $5d$ and $6s$ states are comparable \cite{Ba,La}. Moreover,  triple and higher excitations are
significantly larger for the $5d$ states in comparison to all other states. For example, the difference of the SD and SDpT
values is 398~cm$^{-1}$ for the $5d_{3/2}$ state and only 51~cm$^{-1}$ for the ground state. As a result, some properties of the $5d$ states are  less accurate than the properties of the other states. The scaling procedure mentioned above is used to correct electric-dipole matrix elements involving $5d$ states for missing higher excitations.

\begin {table}
\caption {\label {table2} Recommended values of radiative lifetimes in
nsec). Uncertainties are given in
parenthesis and references are given in brackets.  The values of lifetimes evaluated in the lowest-order DF
approximation are given in column DF to
illustrate the importance of the correlation corrections.
Experimental and other theoretical values are listed in
the two last columns.}
\begin {ruledtabular}
\begin {tabular}{lrrrr}
\multicolumn {1}{c}{Level} &  \multicolumn {1}{c}{DF} &
\multicolumn {1}{c}{Recomm.}& \multicolumn
{1}{c}{Expt.}&
\multicolumn {1}{c}{Other}\\
\hline
 $  6p_{1/2}$&    25.4 &   34.4(1.2)& 34.934(94)~\cite{life-94-6pb}  &   \\
 $  6p_{3/2}$&   22.2  &   30.0(0.7)& 30.460(38)~\cite{life-11-6p}  &  \\

 $  5d_{3/2}$&   600  &   966(34)  & 909(15)~\cite{life-98-5d}  & 976~\cite{life-98-5d}   \\
 $  5d_{5/2}$&   847   &   1351(52) &1281(9) ~\cite{life-98-5d}  & 1363 ~\cite{life-98-5d} \\

 $  7s_{1/2}$&    45.3 &   48.4(0.2)& 49(4) ~\cite{life-77-sdf} & 56 ~\cite{life-77-sdf}\\

 $  7p_{1/2}$&    84   &   152(18)  & 155(4) ~\cite{life-81-7p} & 135 ~\cite{life-81-7p}\\
 $  7p_{3/2}$&    77   &   128(10)  & 133(2)~\cite{life-81-7p}  & 110  ~\cite{life-81-7p}\\

 $  6d_{3/2}$&   153   &     61(2)  & 60.0(2.5)~\cite{life-79-nf}  & 69.9~\cite{life-79-nf} \\
 $  6d_{5/2}$&   138   &     61(2)  & 60.7(2.5)~\cite{life-79-nf}  & 64.5~\cite{life-79-nf}  \\

 $  8s_{1/2}$&    87   &     93(1)  & 87(9) ~\cite{life-77-sdf1} & 104 ~\cite{life-77-sdf1}\\
 $  4f_{7/2}$&    25   &     51(7)  & 40(6)~\cite{life-77-sdf}  &  43~\cite{life-77-sdf}\\
 $  4f_{5/2}$&    24   &     51(7)  & 40(6)~\cite{life-77-sdf}  &  43~\cite{life-77-sdf}    \\

 $  8p_{1/2}$&   201   &   376(16)  & 307(14)~\cite{life-76-np}  &  \\
 $  8p_{3/2}$&   186   &   320(11)  & 274(12)~\cite{life-76-np} &  \\

 $  7d_{3/2}$&   160   &    95(2)   & 89(1)~\cite{life-84-ns-nd}  & 107~\cite{life-84-ns-nd}\\
 $  7d_{5/2}$&   151   &    95(2)   & 89(1)~\cite{life-84-ns-nd}  & 107~\cite{life-84-ns-nd}\\

 $  9s_{1/2}$&   157   &   167(2)   & 159(3)~\cite{life-84-ns-nd}  & 177~\cite{life-84-ns-nd}\\
 $  5f_{7/2}$&    54   &    96(4)   &  97(6)~\cite{life-79-nf}  & 76.8~\cite{life-79-nf}\\
 $  5f_{5/2}$&    53   &    96(4)   &  95(6)~\cite{life-79-nf} &   \\
 $  9p_{1/2}$&   386   &   695(19)  & 575(35)~\cite{life-76-np}  &   \\
 $  9p_{3/2}$&   360   &   606(16)  &  502(22)~\cite{life-76-np} &   \\

 $  8d_{3/2}$&   223   &   153(3)   & 141(2)~\cite{life-84-ns-nd}  & 168~\cite{life-84-ns-nd}\\
 $  8d_{5/2}$&   213   &   153(3)   & 145(3)~\cite{life-84-ns-nd}  &  168~\cite{life-84-ns-nd}\\

 $ 10s_{1/2}$&   262   &   279(3)   & 265(4)~\cite{life-84-ns-nd}  &  293~\cite{life-84-ns-nd}\\
 $  6f_{7/2}$&    97   &   159(3)   &149(8)~\cite{life-79-nf}    &  123.8~\cite{life-79-nf} \\
 $  6f_{5/2}$&    96   &   160(3)   &   &   \\
 $ 10p_{1/2}$&   652   &  1132(29)  & 920(50)~\cite{life-79-10p} &   \\
 $ 10p_{3/2}$&   610   &  1006(26)  & 900(40)~\cite{life-79-10p}  &   \\

 $  9d_{3/2}$&   321   &   235(4)   & 218(3)~\cite{life-84-ns-nd}  & 257~\cite{life-84-ns-nd}\\
 $  9d_{5/2}$&   308   &   237(4)   & 217(4)~\cite{life-84-ns-nd}  & 257~\cite{life-84-ns-nd}\\

 $ 11s_{1/2}$&   408   &   434(4)   & 403(4)~\cite{life-84-ns-nd}  & 455~\cite{life-84-ns-nd}\\
 $  7f_{7/2}$&   158   &   246(3)   &229(15)~\cite{life-79-nf}   &  189.2~\cite{life-79-nf}\\
 $  7f_{5/2}$&   155   &   248(3)   &   &   \\
 $ 11p_{1/2}$&  1012   &  1702(41)  &   &   \\
 $ 11p_{3/2}$&   950   &  1538(37)  &   &   \\

 $ 10d_{3/2}$&   454   &   348(6)   & 315(3)~\cite{life-84-ns-nd}  & 376~\cite{life-84-ns-nd}\\
 $ 10d_{5/2}$&   439   &   350(3)   & 321(4)~\cite{life-84-ns-nd}  & 376~\cite{life-84-ns-nd}\\

 $ 12s_{1/2}$&   602   &   642(6)   & 573(7)~\cite{life-84-ns-nd}  & 668~\cite{life-84-ns-nd}  \\
 $  8f_{7/2}$&   238   &   361(20)  &   &   \\
 $  8f_{5/2}$&   234   &   363(6)   &336(22)~\cite{life-79-nf}   & 274.8~\cite{life-79-nf}  \\
 $ 12p_{1/2}$&  1484   &  2430(55)  &   &   \\
 $ 12p_{3/2}$&  1394   &  2218(52)  &   &   \\

 $ 11d_{3/2}$&   626   &   492(8)   & 417(5)~\cite{life-84-ns-nd}  & 529~\cite{life-84-ns-nd}\\
 $ 11d_{5/2}$&   605   &   496(5)   & 420(7)~\cite{life-84-ns-nd} & 529~\cite{life-84-ns-nd}\\

 $ 13s_{1/2}$&   846   &   901(8)   & 777(8)~\cite{life-84-ns-nd}  &  942~\cite{life-84-ns-nd}\\
 $  9f_{7/2}$&   340   &   506(23)  &473(30)~\cite{life-79-nf}   & 385.1~\cite{life-79-nf}  \\
 $  9f_{5/2}$&   335   &   511(26)  &   &   \\

 $ 12d_{3/2}$&   794   &   663(24)  &566(11)~\cite{life-84-ns-nd}   &  722~\cite{life-84-ns-nd}\\
 $ 12d_{5/2}$&   768   &   661(29)  &586(11)  &  722\\

 $ 14s_{1/2}$&  1033   &  1087(9)   &1017(20)~\cite{life-84-ns-nd}   & 1282~\cite{life-84-ns-nd}\\
 $ 10f_{7/2}$&   419   &   620(27)  &646(35)~\cite{life-79-nf}   & 521.1~\cite{life-79-nf}  \\
 $ 10f_{5/2}$&   414   &   626(29)  &   &   \\
\end{tabular}
\end{ruledtabular}
\end{table}

\subsection{Electric-dipole matrix elements}
\label{matr}
We calculated  the 126 $ns-n'p$ ($n = 6-14$ and $n' = 6-12$),
168 $np-n'd$ ($n = 6-12$ and $n' = 5-12$), and  168 $nd-n'f$ ($n
= 5-12$ and $n' = 4-10$) transitions.
Table~\ref{table1} reports those values
that make significant contributions to
the atomic lifetimes and polarizabilities calculated in the other
sections. The absolute values
are given in all cases in units of $a_0 e$, where $a_0$ is the Bohr radius and $e$ is the elementary charge. More details of the matrix-element calculations, including the lowest-order values, are given in Supplemental Material \cite{SM}.

Unless stated otherwise, we use the conventional system of atomic units,
a.u., in which $e$, the electron mass $m_{\rm e}$, and the reduced
Planck constant $\hbar$ have the numerical value 1, and the electric constant $\epsilon_0$ has the numerical value $1/(4\pi)$.
Dipole polarizabilities $\alpha$ in a.u. have the dimension of volume, and their
numerical values presented here are expressed in units of $a^3_0$. The atomic units
for $\alpha$ can be converted to SI units via
 $\alpha/h$~[Hz/(V/m)$^2$]=2.48832$\times10^{-8}\alpha$~[a.u.], where
 the conversion coefficient is $4\pi \epsilon_0 a^3_0/h$ and the
 Planck constant $h$ is factored out.

The estimated uncertainties of the recommended values are listed in parenthesis.
 The evaluation of the uncertainty of the matrix elements  was
described in detail in \cite{SafSaf11rb,SafSaf12}. It is based on
four different
 all-order calculations mentioned in Section~\ref{Section1}: two \textit{ab initio} all-order calculations with (SDpT)
  and without (SD) the inclusion of the partial triple excitations and two
calculations that included semiempirical estimate of missing high-order
correlation corrections starting from both \textit{ab initio}
calculations.  The spread of these
 four values was used to  estimate uncertainty in the final
results for each transition based on the algorithm  accounting for the importance of
the specific dominant contributions.
The largest  values of the uncertainties in Table~\ref{table1} are
for the  $5d - np$ transitions, ranging from 1.9\% to to 10\% for most cases resulting from larger correlation for the $5d$ states discussed above. The uncertainties are the largest (20\%) for
the $5d - 7p$ transitions.
 Our final
results and their uncertainties
 are used to calculate the recommended values of the  lifetimes and polarizabilities discussed below.

\subsection{Lifetimes}

One of the first lifetime measurements in cesium was
published by Gallagher  \cite{life-67}. Level crossing
measurement of lifetimes in Cs were presented by  Schmieder and
Lurio  \cite{life-70}.  Using a pulsed dye laser and
the method of delayed coincidence the lifetime of the $7p$,
$8p$, and $9p$ levels were
measured by Marek and  Niemax \cite{life-76-np}.
 The cascade
Hanle--effect technique were used by Budos {\it et al.\/}
\cite{life-76-9s} for the lifetime measurements of the $8s$ and $9s$  levels.
Deech {\it et al.\/} \cite{life-77-sd} reported results of the
lifetime measurements  made by time-resolved fluorescence from
$ns$ and $nd_{3/2}$ states of Cs ($n = $8 to 14) over
a range of vapour densities covering the onset of collisional
depopulation.  The same technique was used by Marek
\cite{life-77-sdf} to find out the lifetimes for the $7s$, $5d$, and $4f$ levels.
Radiative lifetimes of the $8s$, $9s$ and $7d$ levels of Cs were
measured by Marek \cite{life-77-sdf1}  employing the method of delayed
coincidences of cascade transitions to
levels that cannot be directly excited by electronic dipole
transitions from ground state.
Alessandretti {\it et al.\/} \cite{life-77-8s} reported
measurement of the $8s$-level lifetime  in Cs vapor, using the two-photon $6s \rightarrow 8s$ transition.
 Marek and  Ryschka \cite{life-79-nf}
presented lifetime measurements of $5f$ - $11f$ levels of Cs
using partially superradiant population.
Ortiz and Campos \cite{life-81-7p} measured
lifetimes of the $7p_{1/2}$   and $7p_{3/2}$
 levels of Cs. Neil and  Atkinson \cite{life-84-ns-nd} reported the
lifetimes of the $ns$  ($n$ = 9-15),  $nd_{3/2}$.
and  $nd_{5/2}$  ($n = 7-12$) levels with 1-2\% accuracy.
Lifetimes were measured by laser-induced fluorescence using
two-photon excitation. Small differences between the lifetimes of
the different fine-structure levels of each $nd$ state have been
observed for the first time \cite{life-84-ns-nd}. Bouchiat {\it et
al.\/} \cite{life-92-5d} reported measurement of the radiative
lifetime of the cesium $5d_{5/2}$ level using pulsed excitation
and delayed probe absorption.
Sasso {\it et al.\/} \cite{life-92-5d} reported measurement of the
radiative lifetimes and quenching of the cesium $5d$ levels.
Measurement of the $5d_{5/2}$ lifetime
was presented by Hoeling {\it et al.\/} \cite{life-96-5d}
and
 by DiBerardino {\it et al.\/} \cite{life-98-5d}.

\begin{table}
\caption{Reduced quadrupole matrix elements $Q$ in a.u. and contributions to quadrupole polarizabilities
of the $6s$ state of cesium in $a_0^5$. Uncertainties are given in
parenthesis. \label{table3}}
\begin{ruledtabular}
\begin{tabular}{lrr}
  \multicolumn{1}{c}{Contr.} &   \multicolumn{1}{c}{$Q$} &
 \multicolumn{1}{c}{$\alpha^{E2}_{0}(6s)$}\\
 [0.3pc]\hline
 $5d_{3/2}$&        33.62& 3421(69)\\
 $6d_{3/2}$&        12.98&  327(21)\\
 $7d_{3/2}$&         8.08&  109(4)\\
 $8d_{3/2}$&         5.53&   48(1)\\
  $(9-26)d_{3/2}$&   &  287(1)\\[0.4pc]
 $5d_{5/2}$&       41.51  & 5183(88)\\
 $6d_{5/2}$&       15.26  &  452(28)\\
 $7d_{5/2}$&        9.64  &  157(6)\\
 $8d_{5/2}$&        6.64 &   70(2) \\
  $(9-26)d_{5/2}$&  &  379(1) \\[0.4pc ]
  CORE     &        &   86(2)\\
    Total    &      &       10521(118) \end{tabular}
\end{ruledtabular}
\end{table}

Precision lifetime measurements of the $6p$ states
in atomic cesium were published in a number of papers by Tanner
{\it et al.\/} \cite{life-92-6p}, Rafac {\it et al.\/}
\cite{life-94-6pa}, Young {\it et al.\/} \cite{life-94-6pb}, and
Rafac {\it et al.\/} \cite{life-99-6p}. The most accurate result
was reported by Rafac {\it et al.\/} \cite{life-94-6pa} with
lifetimes for $6p_{1/2}$ (34.934$\pm$0.094~ns) and $6p_{3/2}$ (30.499$\pm$0.070~ns) states in atomic cesium obtained using
the resonant diode-laser excitation of a fast atomic beam to
produce those measurements.
Recently, lifetime  of the cesium $6p_{3/2}$ state was measured
using ultrafast laser-pulse excitation and ionization. The result of Sell {\it
et al.\/} \cite{life-11-6p}, $\tau (6p_{3/2})= 30.460(38)$~ns  with an uncertainty of 0.12\%, is one of
the most accurate lifetime measurements on record.

We calculated lifetimes of the $(7-14)s$,  $(6-12)p$, $(5-12)d$, and $(4-10)f$ states in
Cs using out final values of the matrix elements  listed in
Table~\ref{table1} and experimental energies from ~\cite{nist-web}.   The uncertainties in
the lifetime values are obtained from the uncertainties in the
matrix elements listed in Table~\ref{table1}. Since experimental energies are very accurate,
the uncertainties in the lifetimes originate from the uncertainties in the matrix elements.
 We also included the lowest-order DF lifetimes
$\tau^{{\rm DF}}$ to illustrate the size of the correlation
effects, which can be estimated as the differences of the final and lowest-order values.
We note that the correlation contributions are large, 5-60\%, being 30-40\% for most states.

Our results are compared with experiment and other theory.
We note that  the theoretical results in the
``$\tau^{{\rm theory}}$'' column are generally quoted in the same papers
as the experimental measurements, with theoretical values
mostly obtained using Coulomb approximation of empirical formulas, which are not expected to be of high
accuracy.

We find
a 20\% difference with the  $8p$ and $9p$ lifetimes measured
by Marek and Niemax \cite{life-76-np} using a pulsed dye
laser and the method of delayed coincidence.
 The same technique was used by Marek
\cite{life-77-sdf} to measure  the lifetimes of the $7s$
 and $4f$  levels.
Our final value is a excellent agreement for the
 $7s$ level, and reasonably agree within the combined uncertainties  for the
$4f$ levels (51(7)~ns vs. 40(6)~ns). Radiative lifetime of the
$8s$ level of Cs was measured by Marek \cite{life-77-sdf1}.
 We confirm a good agreement for the $8s$
level taking uncertainties into account.
Our results are in excellent agreements with the
Ortiz and Campos \cite{life-81-7p}
measurements for the $7p_{1/2}$  and $7p_{3/2}$ lifetimes.
We differ by 5\% - 10\% with the lifetimes of the $ns$
($n = 9-15$),  $nd_{3/2}$. and  $nd_{5/2}$ ($n = 7-12$) levels reported by
 Neil and Atkinson
\cite{life-84-ns-nd}. We note that the uncertainties of
1 - 2\% quoted in \cite{life-84-ns-nd} are most likely underestimated.
Our lifetimes of the $6p_{1/2}$
and $6p_{3/2}$  are in excellent agreements with recent experiment \cite{life-11-6p}
indicating that our uncertainties are overestimated for these levels.

\begin{table}
\caption{\label{table4}  The $\alpha_0$
 scalar and $\alpha_2$ tensor polarizabilities  (in multiples of 1000~a.u.)  in cesium.
 Uncertainties are given in parenthesis. }
\begin{ruledtabular}
\begin{tabular}{llll}
 \multicolumn{1}{c}{nlj} &
 \multicolumn{1}{c}{$\alpha_0$}&
  \multicolumn{1}{c}{nlj} &
 \multicolumn{1}{c}{$\alpha_2$}\\
  [0.3pc]\hline
  $ 7s_{1/2}$&   6.237(42)   &&\\
  $ 8s_{1/2}$&   38.27(26)   &&\\
  $ 9s_{1/2}$&   153.7(1.0)  &&\\
  $10s_{1/2}$&   477.5(3.3)  &&\\
  $11s_{1/2}$&   1246(4)     &&\\
  $12s_{1/2}$&    2868(21)   &&\\
  $13s_{1/2}$&    5817(14)   &&\\  [0.4pc]
 $ 6p_{1/2}$&   1.339(43)    &&\\
  $ 7p_{1/2}$&   29.88(16)    &&\\
  $ 8p_{1/2}$&   223.3(1.4)   &&\\
  $ 9p_{1/2}$&  1021.4(5.6)   &&\\
  $10p_{1/2}$&    3500(14)    &&\\
  $11p_{1/2}$&    9892(37)    &&\\
  $12p_{1/2}$&   24310(100)   &&\\[0.4pc]

  $ 6p_{3/2}$&   1.651(46)   & $ 6p_{3/2}$&   -0.260(11)    \\
  $ 7p_{3/2}$&   37.51(17)   & $ 7p_{3/2}$&   -4.408(50)    \\
  $ 8p_{3/2}$&   284.5(1.7)  & $ 8p_{3/2}$&   -30.57(41)    \\
  $ 9p_{3/2}$&  1312.9(7.0)  & $ 9p_{3/2}$&   -134.7(1.7)    \\
  $10p_{3/2}$&    4525(20)   & $10p_{3/2}$&   -451.1(4.8)    \\
  $11p_{3/2}$&   12836(44)   & $11p_{3/2}$&    -1254(11)    \\
  $12p_{3/2}$&   31630(96)   & $12p_{3/2}$&    -3041(24)    \\[0.4pc]

  $ 5d_{3/2}$&   -0.335(38)  &    $ 5d_{3/2}$&   0.357(25) \\
  $ 6d_{3/2}$&    -5.68(12)  &    $ 6d_{3/2}$&   8.749(78) \\
  $ 7d_{3/2}$&   -66.79(1.0) &    $ 7d_{3/2}$&   71.08(75) \\
  $ 8d_{3/2}$&   -369.3(5.8  &    $ 8d_{3/2}$&  338.6(3.1)  \\
  $ 9d_{3/2}$&    -1405(20)  &    $ 9d_{3/2}$&     1190(8)   \\
  $10d_{3/2}$&    -4242(60)  &    $10d_{3/2}$&    3419(22) \\
  $11d_{3/2}$&   -10926(149) &    $11d_{3/2}$&    8511(54)      \\
  $12d_{3/2}$&   -25092(783) &    $12d_{3/2}$&   18734(158)    \\[0.4pc]

  $ 5d_{5/2}$&   -0.439(42)  &   $ 5d_{5/2}$&   0.677(34)  \\
  $ 6d_{5/2}$&    -8.38(13)  &   $ 6d_{5/2}$&   17.30(11)  \\
  $ 7d_{5/2}$&    -88.9(1.3) &   $ 7d_{5/2}$&   141.7(1.1)  \\
  $ 8d_{5/2}$&   -475.8(6.3) &   $ 8d_{5/2}$&   678.1(4.9)  \\
  $ 9d_{5/2}$&    -1781(22)  &   $ 9d_{5/2}$&    2388(14)  \\
  $10d_{5/2}$&    -5324(59)  &   $10d_{5/2}$&    6871(35)  \\
  $11d_{5/2}$&   -13615(136) &   $11d_{5/2}$&   17111(69)      \\
  $12d_{5/2}$&   -31487(802  &   $12d_{5/2}$&   38424(291)     \\
\end{tabular}
\end{ruledtabular}
\end{table}

\begin{figure*}[tbp]
\includegraphics[scale=0.31]{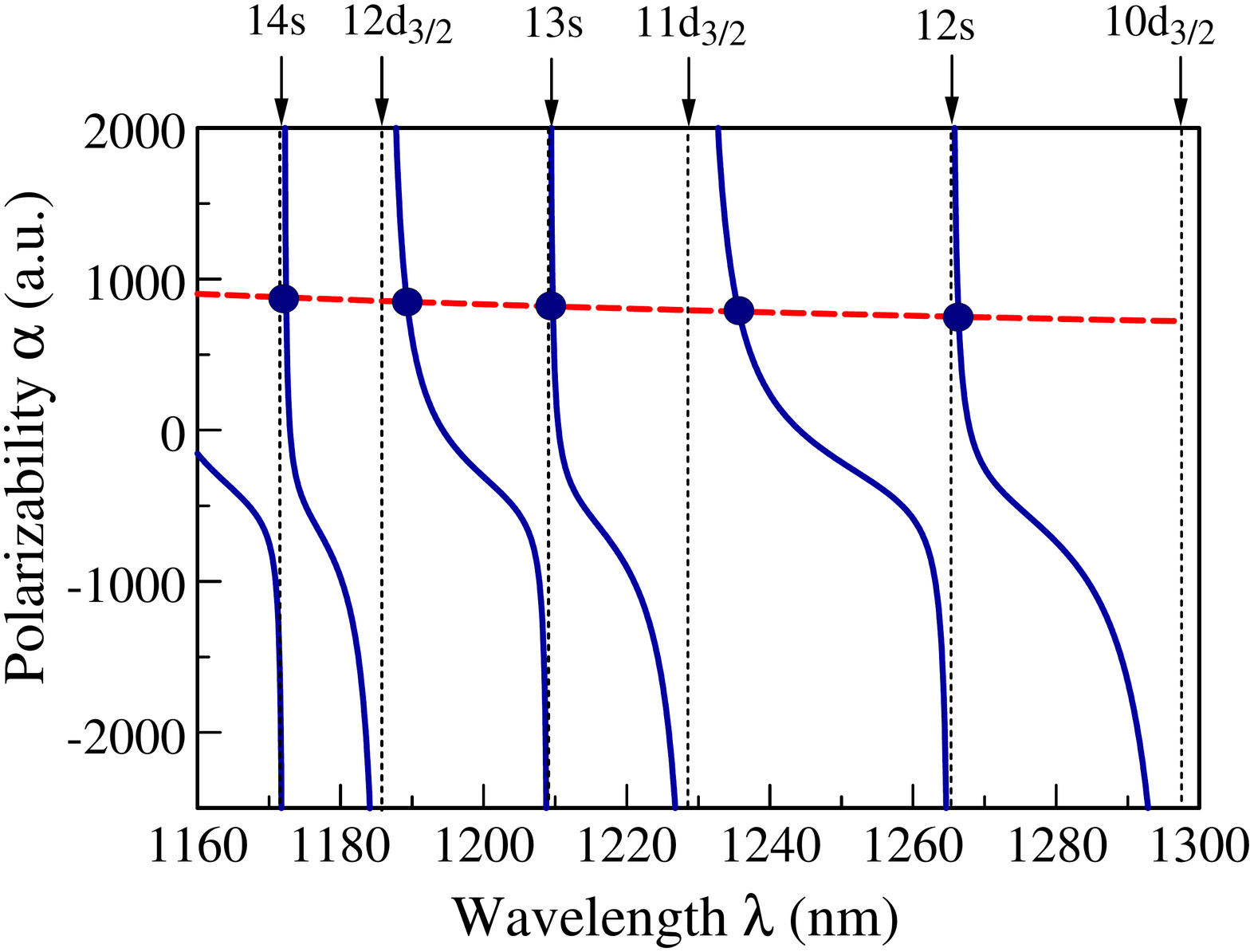} \hspace{0.1in}
\includegraphics[scale=0.31]{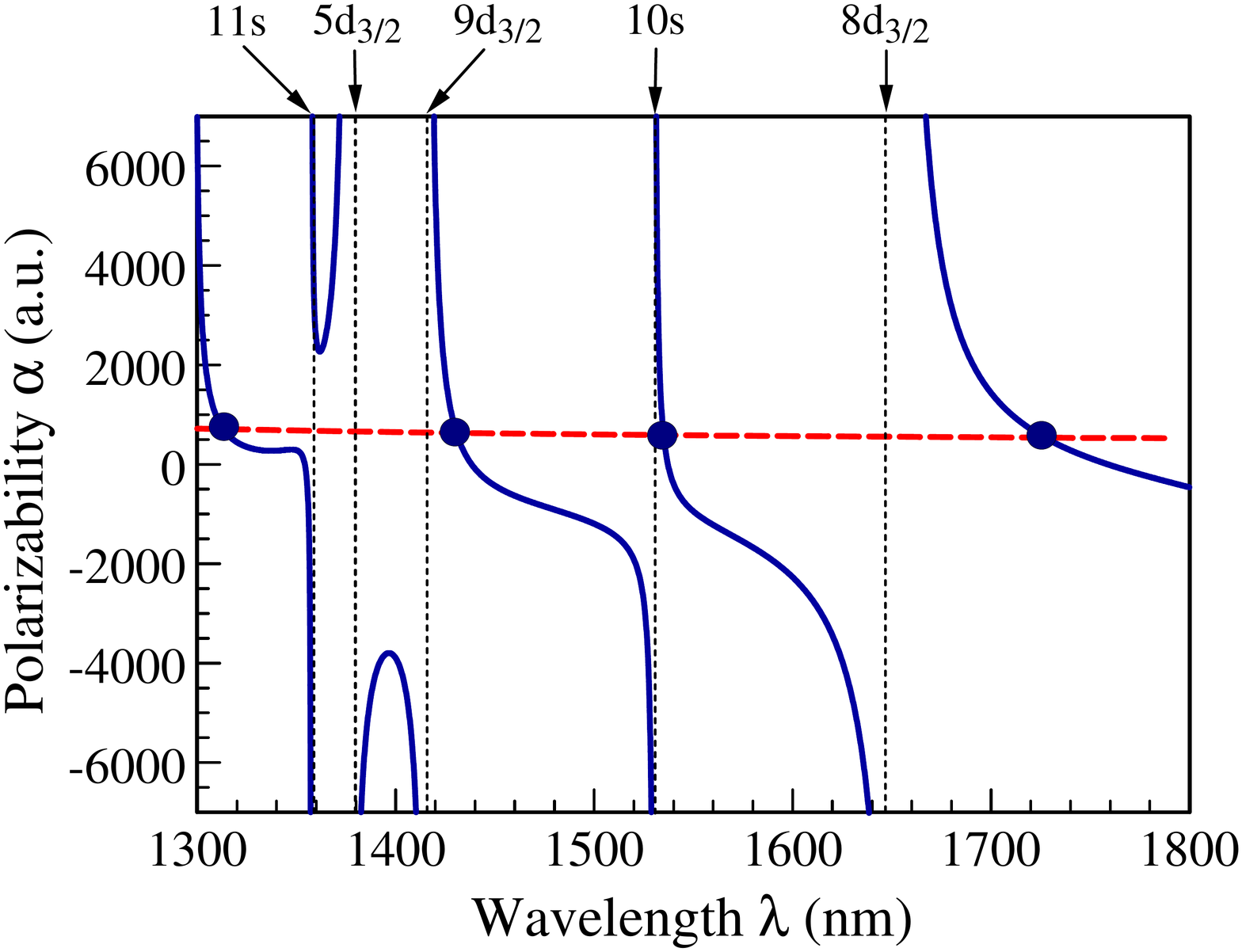}
  \caption{(Color online) The frequency-dependent polarizabilities
of the Cs $6s$ and $7p_{1/2}$ states. The magic wavelengths are marked
with circles. The approximate positions of the $7p_{1/2}-nl$
resonances are indicated by vertical lines with small
arrows on top of the graph.}\label{fig2ab}
\end{figure*}

\begin{figure}[tbp]
\includegraphics[scale=0.3]{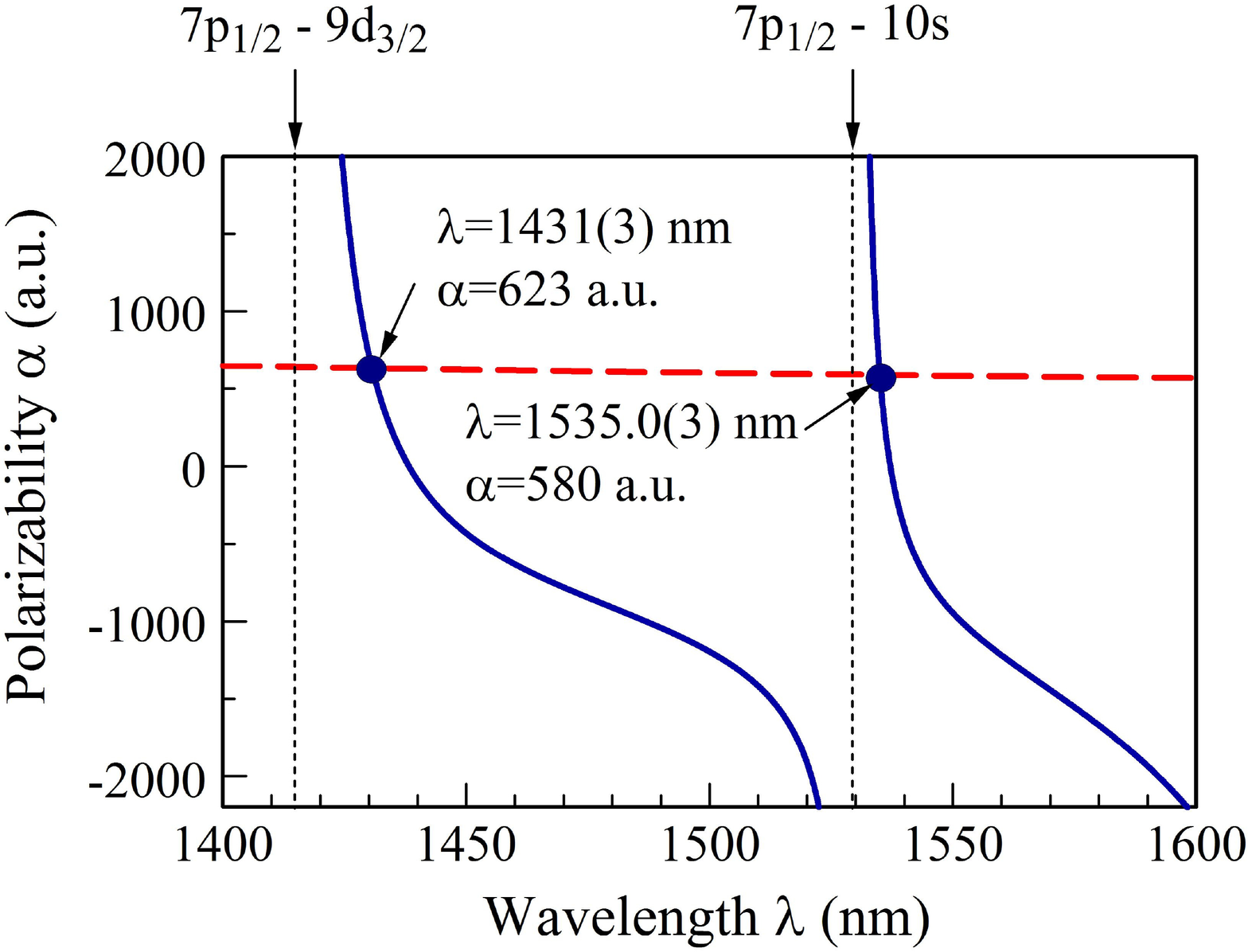}
  \caption{(Color online) The frequency-dependent polarizabilities
of the Cs $6s$ and $7p_{1/2}$ states. The magic wavelengths are marked
with circles and arrows. The approximate positions of the $7p_{1/2}-9d_{3/2}$
and $7p_{1/2}-10s$ resonances are indicated by vertical lines with small
arrows on top of the graph.}\label{fig1a}
\end{figure}

\begin{figure}[tbp]
\includegraphics[scale=0.31]{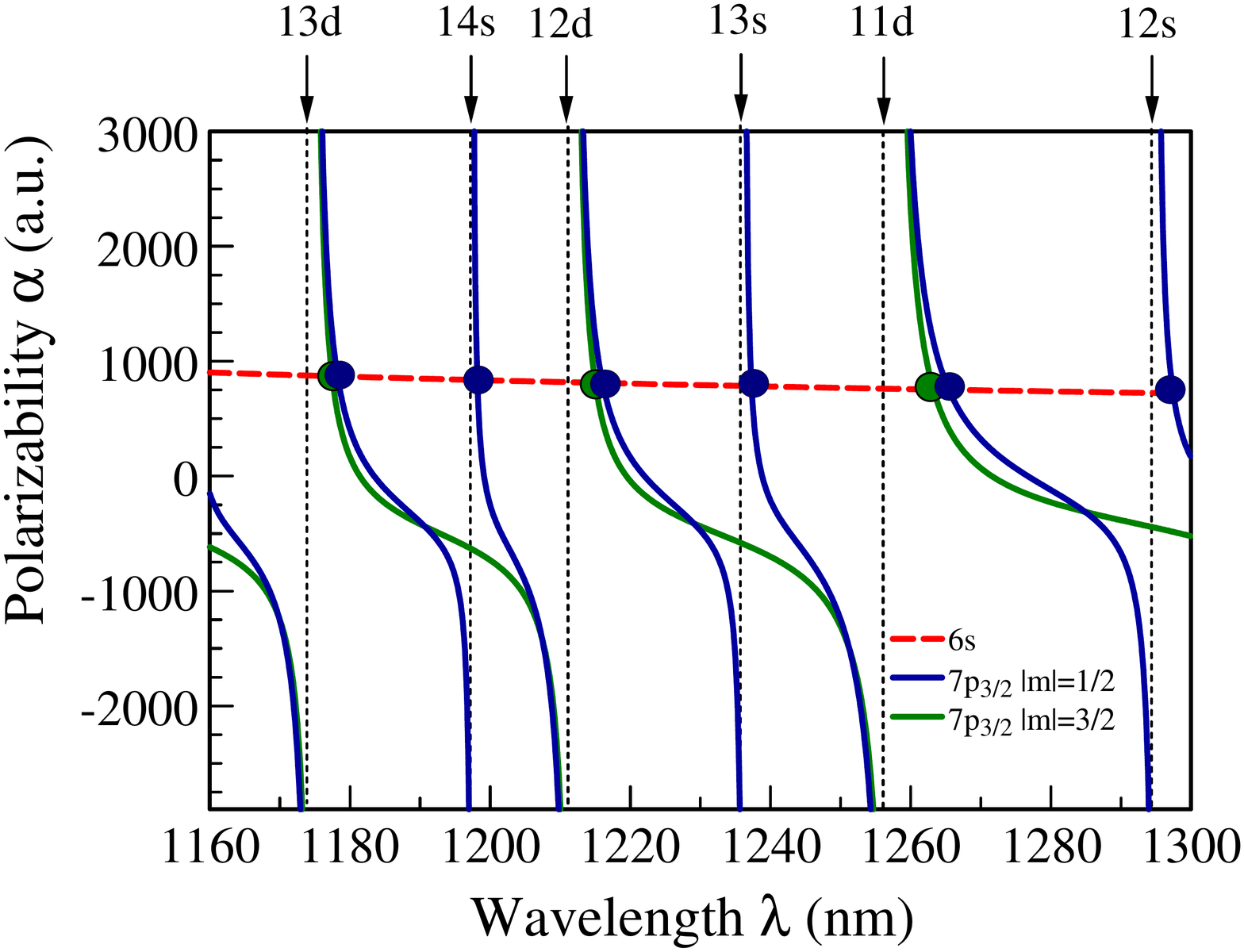}
  \caption{(Color online) The frequency-dependent polarizabilities
of the Cs $6s$ and $7p_{3/2}$ states. The magic wavelengths are marked
with circles. The approximate positions of the $7p_{3/2}-nl$
resonances are indicated by vertical lines with small
arrows on top of the graph.}\label{fig2c}
\end{figure}

\begin{figure*}[tbp]
\includegraphics[scale=0.31]{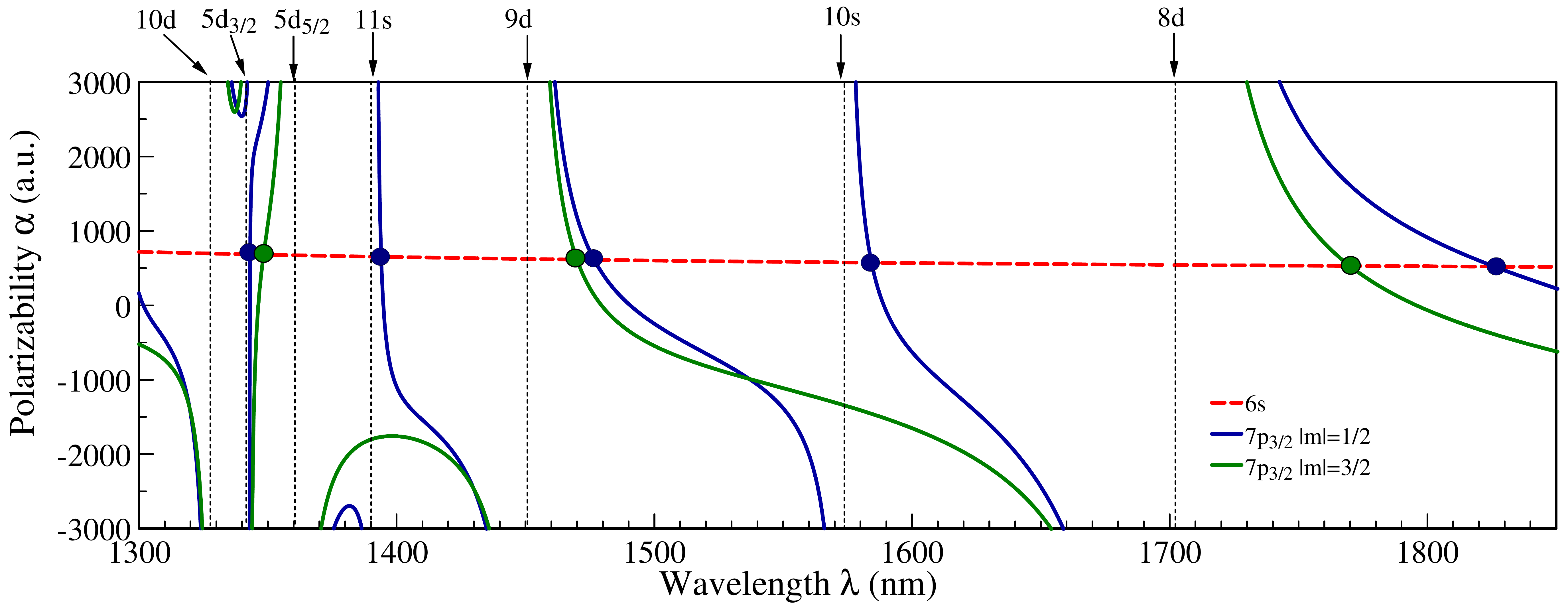}
  \caption{(Color online) The frequency-dependent polarizabilities
of the Cs $6s$ and $7p_{3/2}$ states. The magic wavelengths are marked
with circles and arrows. The approximate positions of the $7p_{3/2}-nl$
resonances are indicated by vertical lines with small
arrows on top of the graph.}\label{fig2d}
\end{figure*}

\section{ Static quadrupole polarizabilities  of  the $6s$ state}

The static multipole polarizability $\alpha^{Ek}$ of Cs
 in its $6s$  state can be separated into two terms; a
dominant first term
 from intermediate valence-excited states, and a smaller
  second term from the core-excited  states.
  The second term is the lesser of these and is  evaluated here in the
random-phase approximation \cite{RPA}. The dominant valence
contribution  is calculated using the sum-over-state approach
\begin{equation}
\alpha^{Ek}_v = \frac{1}{2k+1}\sum_{n} \frac{|\langle nl_j\| r^k
C_{kq} \|6s\rangle|^2}{E_{nlj}-E_{6s}}, \label{alpha0}
\end{equation}
where  $C_{kq}(\hat{r})$ is a normalized spherical harmonic and
where $nl$ is $np$, $nd$, and $nf$ for $k$ = 1, 2, and
3, respectively \cite{multipol}. Here we dicuss the  the
quadrupole ($k =  2$) polarizabilities.

 We use recommended energies from \cite{nist-web} and our
 final quadrupole matrix elements to evaluate terms in the
sum with $n \leq 13$, and we use theoretical SD energies and
matrix elements to evaluate terms with $13\leq n \leq 26$. The
remaining contributions to $\alpha^{E2}$ from orbitals with $27
\leq n \leq 70$ are evaluated in the random-phase approximation (RPA). We find that this contribution is
negligible.
The uncertainties in the polarizability
contributions are obtained from the uncertainties in the
corresponding matrix elements. The final values for the quadrupole
 matrix elements and their uncertainties are
determined using the same procedure as for the dipole matrix
elements.

 Contributions to  the quadrupole
polarizability of the $6s$ ground state are presented in
Table~\ref{table3}.
While the first two terms in the sum-over-states
for the electric dipole polarizability contribute 99.5\%, the first two terms in the
sum-over-states for $\alpha^{E2}$  contribute 82.4\%. The first
eight terms gives 93.6\%.
 The remaining 6.4\% of $\alpha^{E2}$ contributions are from
the $(9-26)nd$ states. Single-photon laser excitation of the $6s-5d$ transition has been used in Cs spectroscopy \cite{PhysRevA.35.4650}, and the transition rate can be calculated from data in Table~\ref{table3}.

\begin{figure}[tbp]
\includegraphics[scale=0.3]{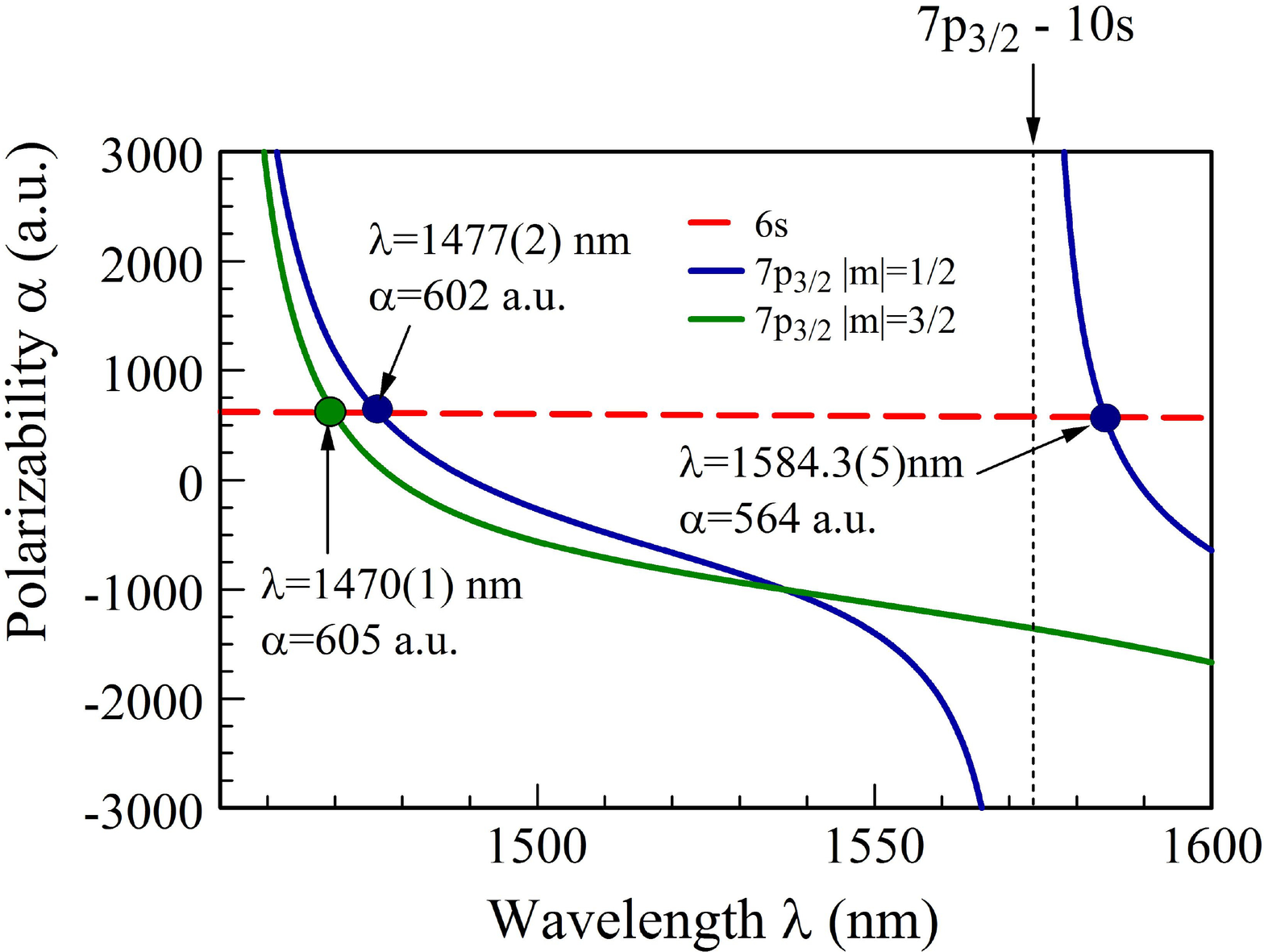}
  \caption{(Color online) The frequency-dependent polarizabilities
of the Cs $6s$ and $7p_{3/2}$ states. The magic wavelengths are marked
with circles and arrows. The approximate position of the
and $7p_{3/2}-10s$ resonance is indicated by vertical line with small
arrows on top of the graph.}\label{fig1b}
\end{figure}

\section{ Scalar and tensor  polarizabilities for excited states of cesium}

The frequency-dependent scalar polarizability, $\alpha(\omega)$, of an
alkali-metal atom in the state $v$ may be separated into a
contribution from the ionic core, $\alpha_{\rm{core}}$, a core
polarizability modification due to the valence electron,
$\alpha_{vc}$, and a contribution from the valence electron,
$\alpha^v(\omega)$. We find scalar
Cs$^+$ ionic core polarizability, calculated
in random-phase approximation (RPA)
to be  15.84~$a_0^{3}$, which is consistent with other data (see Table 4 of Ref.  \cite{mitroy-10}).
A counter term $\alpha _{\rm vc}$
compensates for Pauli principle violating core-valence excitation from the core to the valence shell. It is small, $\alpha _{\rm vc} = -0.673$~a.u.
for the $6s$ state of Cs.
Since the core is isotropic, it makes no contribution to tensor polarizabilities.

 The valence contribution to frequency-dependent scalar $\alpha_0$ and
 tensor $\alpha_2$ polarizabilities is
evaluated as the sum over intermediate $k$ states allowed by the
electric-dipole selection rules~\cite{mitroy-10}
\begin{eqnarray}
    \alpha_{0}^v(\omega)&=&\frac{2}{3(2j_v+1)}\sum_k\frac{{\left\langle k\left\|D\right\|v\right\rangle}^2(E_k-E_v)}{     (E_k-E_v)^2-\omega^2}, \label{eq-1} \nonumber \\
    \alpha_{2}^v(\omega)&=&-4C\sum_k(-1)^{j_v+j_k+1}
            \left\{
                    \begin{array}{ccc}
                    j_v & 1 & j_k \\
                    1 & j_v & 2 \\
                    \end{array}
            \right\} \nonumber \\
      & &\times \frac{{\left\langle
            k\left\|D\right\|v\right\rangle}^2(E_k-E_v)}{
            (E_k-E_v)^2-\omega^2} \label{eq-pol},
\end{eqnarray}
             where $C$ is given by
\begin{equation}
            C =
                \left(\frac{5j_v(2j_v-1)}{6(j_v+1)(2j_v+1)(2j_v+3)}\right)^{1/2} \nonumber
\end{equation}
In the equations above,
$\omega$ is assumed to be at least several linewidths off
resonance with the corresponding transitions and ${\left\langle k\left\|D\right\|v\right\rangle}$ are the
reduced electric-dipole matrix elements. Linear polarization
is assumed in all calculation.  To calculate static polarizabilities, we take $\omega=0$.
The excited state polarizability calculations are carried out in
the same way as the calculations of the multipole polarizabilities
discussed in the previous section.

Contributions to the polarizabilities  of the $6p_{1/2}$,
$6p_{3/2}$ levels,
$5d_{3/2}$ and $5d_{5/2}$ state of cesium are given in the Supplemental material \cite{SM}.
In Table~\ref{table5}, we list the
$\alpha_0$ scalar and $\alpha_2$ tensor polarizabilities  (in multiples of 1000~a.u.)  in cesium.
 Uncertainties are given in parenthesis.

The largest (86.6\%) contribution to the $\alpha_{0}(6p_{1/2})$ value
arises from the the $6p_{1/2}- 5d_{3/2}$ transition. The contribution of the
$6s$ and $7s$ states in the $\alpha_{0}(6p_{1/2})$
value nearly cancel each other.
Some cancellations are also observed in the breakdown of the $5d_{3/2}$ polarizability.
We find that highly-excited $(9-26)f$
states contribute significantly, to the $5d_{3/2}$ and $5d_{5/2}$ polarizabilities, 14\% and 11\%,
respectively.

We list the scalar
polarizabilities of the $(7 - 13)s$, $(6-12)np$, and  $(5 - 12)d$, and tensor polarizabilities of the
 $(6-12)np_{3/2}$  and $(5-12)d$  states
 in Table~\ref{table4}. Uncertainties are given in
 parenthesis.
 Comparison with
theoretical results from van Wijngaarden and  Li \cite{pol-94a},
Iskrenova-Tchoukova {\it et al.\/} \cite{pol-cs-07}, and Mitroy
{\it et al.\/} \cite{mitroy-10} are given in the Supplemental Material \cite{SM}.
 Results in the review paper \cite{mitroy-10} are taken  from
 paper \cite{pol-cs-07}.
The calculations of Ref.~\cite{pol-cs-07} were also obtained using the single-double  all-order method.
 In the present work,
we treat higher-excited states more accurately, carrying all-order calculations
up to $n=26$ instead of $n=12$. As we noted above, higher-excited states are
particulary important for the $nd$ polarizabilities so the only significant differences with results of
~\cite{pol-cs-07,mitroy-10} occur for the $5d$ polarizabilities.

The scalar and tensor polarizabilities in \cite{pol-94a}
 were evaluated using the
Coulomb approximation. The expected
scaling of polarizabilities as $(n^*)^7$, where $n^*$ is the effective
principal quantum number, was found to hold well for the higher
excited states. Our values for the $n$ = 11 and 12 polarizabilities agree to
 1\% with \cite{pol-94a}.
\begin{table} \caption[]{Magic wavelengths in nm for the $6s-7p$
transitions in Cs in the 1160-1800~nm wavelength range. The
corresponding polarizabilities at magic wavelengths  are given in a.u. The resonances
near the magic wavelengths are listed in the first column.}
\label{table5}
\begin{ruledtabular}
\begin{tabular}{lrr}
   \multicolumn{1}{c}{Resonance} &
   \multicolumn{1}{c}{$ \lambda_\textrm{magic}$} &
  \multicolumn{1}{c}{$\alpha(\lambda_\textrm{magic})$} \\
\hline
   \multicolumn{3}{c}{$6s-7p_{1/2}$ transition}\\
  $7p_{1/2} - 14s      $& 1172.40(3)  &   866   \\ $7p_{1/2} - 12d_{3/2}$& 1189.3(4)   &   838  \\
  $7p_{1/2} - 13s      $& 1209.68(4)  &   807  \\
  $7p_{1/2} - 11d_{3/2}$& 1235.7(5)   &   774  \\
  $7p_{1/2} - 12s      $& 1266.4(1)   &   740 \\
  $7p_{1/2} - 10d_{3/2}$& 1313(6)     &   698 \\
  $7p_{1/2} - 9d_{3/2} $& 1431(3)     &   623  \\
  $7p_{1/2} - 10s      $& 1535.0(3)   &   580  \\
  $7p_{1/2} - 8d_{3/2} $& 1727(5)     &   530  \\[0.2pc]
   \multicolumn{3}{c}{$6s-7p_{3/2}$, $|m|=1/2$ transition}\\
  $7p_{3/2} - 13d_{3/2}$&1178.3(4)  &    856    \\
  $7p_{3/2} - 14s      $&1198.18(4) &    824    \\
  $7p_{3/2} - 12d_{3/2}$&1216.1(4)  &    799    \\
  $7p_{3/2} - 13s      $&1237.31(7) &    772    \\
  $7p_{3/2} - 11d_{3/2}$&1265.5(8)  &    741    \\
  $7p_{3/2} - 12s      $&1297.5(4)  &    711    \\
  $7p_{3/2} - 5d_{3/2}$ &11343(2)   &    675    \\
  $7p_{3/2} - 11s      $&1394.0(3)  &    643    \\
  $7p_{3/2} - 9d_{3/2} $&1477(2)    &    602    \\
  $7p_{3/2} - 10s      $&1584.3(5)  &    564     \\
  $7p_{3/2} -  8d_{3/2}$&1827(6)    &    512     \\[0.2pc]
\multicolumn{3}{c}{$6s-7p_{3/2}$, $|m|=3/2$ transition}\\
  $7p_{3/2} - 13d_{3/2}$& 1177.5(4) &  857     \\
  $7p_{3/2} - 12d_{3/2}$& 1215.0(4) &  800    \\
  $7p_{3/2} - 11d_{5/2}$& 1263.3(5) &  743     \\
   $7p_{3/2} - 5d_{3/2}$& 1348(4)   &  671      \\
  $7p_{3/2} - 9d_{5/2} $& 1470(1)   &  605      \\
  $7p_{3/2} - 8d_{5/2} $& 1770(3)   &  521      \end{tabular}
\end{ruledtabular}
\end{table}
\section{Magic wavelengths}
\label{sec3}
Magic wavelengths for $D_1$ and $D_2$ lines in alkali-metal atoms were recently
investigated in
Refs.~\cite{AroSafCla07,magic-08,magic-10,magic-11}.  Flambaum {\it et al.\/} \cite{magic-08}
considered magic conditions for the ground state hyperfine clock transitions of
cesium and rubidium atoms which used as the primary and the
secondary frequency standards. The theory of magic optical traps for
Zeeman-insensitive clock transitions in alkali-metal atoms was
developed by Derevianko \cite{magic-10}. Zhang  {\it et al.\/}
\cite{magic-11} proposed blue-detuned optical traps that were
suitable for trapping of both ground-state and Rydberg excited
atoms.

Several magic wavelengths were calculated for the $6s-6p_{1/2}$
and $6s-7p_{3/2}$ transitions in Cs in Ref.~\cite{AroSafCla07}
using the all-order approach.
  In this work, we present
several other magic wavelengths for for the $6s-7p_{1/2}$ and
$6s-7p_{3/2}$ transitions in Cs.

The magic wavelength $\lambda_{\rm{magic}}$ is defined as the
wavelength
 for which the frequency-dependent  polarizabilities of  two atomic states
  are the same, leading to a
vanishing ac Stark shift for that transition. To determine magic wavelengths
for the $6s-7p$
transition, one calculates $\alpha_{6s}(\lambda)$ and
$\alpha_{7p}(\lambda)$ polarizabilities and
finds  the wavelengths  at
which two respective curves intersect.
All calculations are carried out for linear polarization.

The frequency-dependent polarizabilities are calculated the same way as the
static polarizabilities, but setting $\omega \neq 0$.
The dependence of the core polarizability on the frequency is negligible for the infrared frequencies of
interests for this work. Therefore, we use the RPA static numbers for the ionic core and
$\alpha_{vc}$ terms.

The total polarizability is given by
$$
\alpha=\alpha_0+\alpha_2 \frac{3m^2-j(j+1)}{j(2j-1)},
$$
where $j$ is the total angular momentum and  $m$ is corresponding magnetic quantum number.
 The total polarizability for the $7p_{3/2}$ states is
given by
 $$\alpha=\alpha_0-\alpha_2$$ for $m=\pm 1/2$
and $$\alpha=\alpha_0+\alpha_2$$  for the $m=\pm 3/2$ case.
Therefore, the total polarizability of the $7p_{3/2}$ state
depends
 upon its $m$ quantum number and
  the magic wavelengths needs to be determined separately for the cases
  with $m=\pm 1/2$ and
$m=\pm 3/2$  for the $6s-7p_{3/2}$ transitions, owing to the
presence of
 the tensor contribution to the total
polarizability of the $7p_{3/2}$ state.
There is no tensor contribution to the polarizability of the $7p_{1/2}$ state.
To determine the uncertainty in the values of magic wavelengths, we first determine
the uncertainties in the polarizability values at the magic wavelengths.
Then, the uncertainties in the values of magic wavelengths are determined as the maximum
 differences between
 the central value and the crossings of the
$\alpha_{6s} \pm \delta \alpha_{6s}$ and $\alpha_{7p} \pm \delta
\alpha_{7p}$
 curves, where the  $\delta \alpha$
are the uncertainties in the corresponding $6s$ and $7p$
polarizabilities.

Our magic wavelength results are given by Figs.~\ref{fig2ab}, \ref{fig1a}, \ref{fig2c}, \ref{fig2d}, \ref{fig1b},
and Table~\ref{table5}.
 The frequency-dependent polarizabilities of the $6s$ and
$7p_{1/2}$ states for $\lambda$ =1160~nm~$-$~1800~nm are plotted in
Fig.~\ref{fig2ab}. The magic wavelengths occur between the resonances corresponding
to the $7s_{1/2}-nl$ transitions since the $6s$ polarizability curve has no resonances
in this region and is nearly flat. Magic wavelengths are indicated by filled circles.
The approximate positions of the  $7s_{1/2}-nl$ resonances are indicated by the lines with small
arrows on top of the graph, together with the corresponding
$nl$ label.  The  $\lambda$ =1160~nm~$-$~1800~nm region contains resonances with $nl=5d,8d-12d$ and
$nl=10s-14s$. Resonant wavelengths are listed in the last table of the Supplemental Material ~\cite{SM}.
 The $5d$ energy levels are below the $7p$ energy levels, while all of the other levels are
above the $7p$, leading to interesting features of the $7p$ polarizability curves near the $7p-5d$
resonances.
Due to particular experimental interest in the magic wavelength in the region nearly 1550~nm
due to availability of the corresponding laser, we show  more detailed plot of the
frequency-dependent polarizabilities of the $6s$ and
$7p_{1/2}$ states in the $\lambda$ = 1440~$-$~1600~nm  region in Fig.~\ref{fig1a}.
The  numerical values of these magic wavelengths are given in Table~\ref{table5}.

 The frequency-dependent polarizabilities of the $6s$ and
$7p_{3/2}$ states for $\lambda$ =1160~nm~$-$~1850~nm are plotted in
Figs.~\ref{fig2c} and \ref{fig2d}. The  numerical values of these magic wavelengths are given in Table~\ref{table5}. A detailed plot of the  $\lambda$ = 1440~$-$~1600~nm  region is
shown in  Fig.~\ref{fig1b}. With the exception of the $7p_{3/2}-5d$ case, $7p_{3/2}-nd_{3/2}$ and $7p_{3/2}-nd_{5/2}$
resonances are too close together to show by separate lines on the plots
due to small difference in the $nd_{3/2}$ and $nd_{5/2}$ energies for large $n$.
Therefore, we indicate both $7p_{3/2}-nd_{3/2}$ and $7p_{3/2}-nd_{5/2}$ resonances by a single
vertical line in Figs.~\ref{fig2c}, \ref{fig2d}, and ~\ref{fig1b} with the  ``$nd$'' label on the top.
While there will be additional magic wavelengths in between the $7p_{3/2}-nd_{3/2}$ and $7p_{3/2}-nd_{5/2}$
resonances, we expect them to be unpractical to use in the experiment due to very strong dependence of polarizabilities
on the wavelengths in these cases. Therefore, we omit such magic wavelengths in Table~\ref{table5} and corresponding figures.

In summary, we carried out a systematic study of Cs atomic properties using all-order methods.
Several calculations are carried out to evaluate uncertainties of the final results.
Cs properties are needed for interpretation of the current experiments as well as
planning of future experimental studies.

This research was performed under the sponsorship of the
U.S. Department of Commerce, National Institute of Standards
and Technology, and was supported by the National Science
Foundation via  the Physics Frontiers Center at the Joint Quantum Institute.

\end{document}